\UseRawInputEncoding
\documentclass[draftcls,peerreview]{IEEEtran}

\usepackage{graphicx} 
\usepackage[utf8]{inputenc}
\usepackage[T1]{fontenc}
\usepackage[english]{babel}
\usepackage[unicode=true]{hyperref}
\usepackage{amsmath}
\usepackage{amssymb}
\usepackage{mathtools}
\usepackage[caption=false, font=footnotesize]{subfig}

\usepackage{csquotes}
\usepackage{microtype}

\usepackage{orcidlink}

\makeatletter
\renewcommand{\fnum@figure}{Fig. \thefigure}
\makeatother

\begin{document}
\bstctlcite{IEEEexample:BSTcontrol}

\title{Speech-dependent Data Augmentation for Own~Voice Reconstruction with Hearable Microphones in Noisy Environments}

\author{
Mattes Ohlenbusch,
Christian Rollwage, 
Simon Doclo

\thanks{
The Oldenburg Branch for Hearing, Speech and Audio Technology HSA is funded in the program »Vorab« by the Lower Saxony Ministry of Science and Culture (MWK) and the Volkswagen Foundation for its further development.
This work was partly funded by the German Ministry of Science and Education BMBF FK 16SV8811 and the Deutsche Forschungsgemeinschaft (DFG, German Research Foundation) - Project ID 352015383 – SFB 1330 C1.
}

\thanks{%
M.~Ohlenbusch and C.~Rollwage are with the Fraunhofer Institute for Digital Media Technology IDMT, Oldenburg Branch for Hearing, Speech and Audio Technology HSA, Germany (email: mattes.ohlenbusch@idmt.fraunhofer.de; christian.rollwage@idmt.fraunhofer.de).}%
\thanks{%
S.~Doclo is with the Fraunhofer Institute for Digital Media Technology IDMT, Oldenburg Branch for Hearing, Speech and Audio Technology HSA, Germany, and the Department of Medical Physics and Acoustics and Cluster of Excellence Hearing4all, Carl von Ossietzky Universität Oldenburg, Germany (email: simon.doclo@uni-oldenburg.de).}
\thanks{Manuscript received xxx; revised xxx.}%
}

\markboth{Journal of \LaTeX\ Class Files,~Vol.~14, No.~8, August~2021}%
{Shell \MakeLowercase{\textit{et al.}}: A Sample Article Using IEEEtran.cls for IEEE Journals}
\maketitle

\begin{abstract}
Own voice pickup for hearables in noisy environments benefits from using both an outer and an in-ear microphone
outside and inside the occluded ear.
Due to environmental noise recorded at both microphones, and amplification of the own voice at low frequencies and band-limitation at the in-ear microphone, an own voice reconstruction system is needed to enable communication. 
A large amount of own voice signals is required to train a supervised deep learning-based own voice reconstruction system.
Training data can either be obtained by recording a large amount of own voice signals of different talkers with a specific device, which is costly, or through augmentation of available speech data. 
Own voice signals can be simulated by assuming a linear time-invariant relative transfer function between hearable microphones for each phoneme, referred to as own voice transfer characteristics. 
In this paper, we propose data augmentation techniques for training an own voice reconstruction system based on speech-dependent models of own voice transfer characteristics between hearable microphones. 
The proposed techniques use few recorded own voice signals to estimate transfer characteristics and can then be used to simulate a large amount of own voice signals based on single-channel speech signals.
Experimental results show that the proposed speech-dependent individual data augmentation technique leads to better performance compared to other data augmentation techniques or compared to training only on the available recorded own voice signals, and additional fine-tuning on the available recorded signals can improve performance further. 
\end{abstract}

\begin{IEEEkeywords}
Data augmentation, hearables, multi-microphone speech enhancement, own voice reconstruction, voice pickup.
\end{IEEEkeywords}

\begin{figure}[ht] 
    \centering
    \includegraphics[width=0.6\columnwidth,page=2]{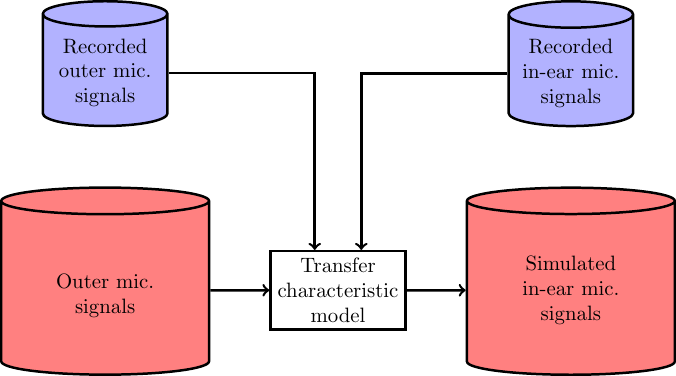} 
    \caption{
	Data augmentation with own voice transfer characteristic models. 
	The models are estimated based on recorded hearable signals at an outer and an in-ear microphone. 
	During data augmentation, in-ear signals are simulated with the own voice transfer characteristic models using a large amount of outer microphone signals.
	} 
    \label{fig:data_augmentation} 
\end{figure}

\section{Introduction}
%
\IEEEPARstart{C}{ommunication} is often difficult in noisy work environments, such as in industrial production or surgery rooms. 
Own voice pickup can improve communication considerably, e.g., for transmitting the own voice of the user to a mobile phone or another hearable~\cite{nordholm_assistive_2015}. 
Due to safety equipment such as protective helmets or breathing masks, headset-like microphones in front of the talkers mouth are often not practical. 
In such cases, in-the-ear hearables occluding the ear canal with integrated microphones still offer communication benefits. 
Both a microphone at the outer face of the hearable (outer microphone) and a microphone inside the occluded ear canal (in-ear microphone) are useful to pick up the own voice of the hearable user.
When recording the own voice in a noisy environment, an outer microphone and an in-ear microphone also pick up additive environmental noise.
Since the hearable is occluding the ear canal, in this scenario an in-ear microphone may be particularly beneficial for picking up the own voice since environmental noise is attenuated.
However, compared to own voice recorded at an outer microphone, own voice recorded at an in-ear microphone is also subject to amplification at low frequencies (below ca.\,1\,kHz) and strong attenuation at higher frequencies (above ca.\,2\,kHz), leading to a limited bandwidth.
In-ear microphones also record body-produced noise~\cite{bouserhal_-ear_2019}.
The own voice recorded at an in-ear microphone mostly consists of body-conducted speech.
The ratio between the airborne and the body-conducted component of own voice recorded at an in-ear microphone depends on speech content~\cite{reinfeldt_hearing_2010,saint-gaudens_towards_2022}, possibly due to mouth movements during articulation~\cite{richard_effect_2023} or the speech originating from different places of excitation~\cite{porschmann_influences_2000}.
This ratio is also subject to individual differences due to anatomic factors such as residual ear canal volume and shape~\cite{stenfelt_model_2007,vogl_individualized_2019}. 
In this paper, the speech-dependent, individual relation between own voice recorded at an outer microphone and recorded at an in-ear microphone is referred to as own voice transfer characteristics. 
These transfer characteristics are assumed to consist of a linear, time-invariant relative transfer function for each phoneme.
Due to additive noise and the influence of the own voice transfer characteristics, neither an outer microphone nor an in-ear microphone can be used for recording the own voice without reconstructing it from noisy microphone signals.
An own voice reconstruction system using both an outer microphone and an in-ear microphone needs to jointly perform noise reduction, bandwidth extension and equalization in order to account for environmental noise and own voice transfer characteristics.

%
Previous approaches based on classical signal processing for reconstructing own voice recorded by in-ear microphones or body-conduction sensors have been proposed that either rely on equalization filter design~\cite{kondo_equalization_2006}, linear prediction analysis and synthesis~\cite{rahman_intelligibility_2011}, or statistical modeling~\cite{shin_survey_2012, shin_priori_2015}.
In~\cite{bouserhal_-ear_2017}, nonlinear bandwidth extension methods have been applied to recorded in-ear own voice signals for reconstructing high-frequency own voice content.
Many recent approaches to own voice reconstruction rely on deep learning, where large amounts of own voice signals are required as training data. 
Deep learning-based own voice reconstruction systems using either in-ear microphones or body-conduction microphones with similar characteristics have been proposed, where the systems are trained directly using device-specific datasets of recorded own voice signals~\cite{liu_bone-conducted_2018, park_speech_2019,  liu_multichannel_2020, yu_time-domain_2020, wang_multi-modal_2022, wang_attention-based_2022, li_enabling_2023, ohlenbusch_multi-microphone_2024, li_restoration_2024, li_two-stage_2024}. 
Any system trained on a dataset of recorded own voice signals, e.g., the ESMB corpus~\cite{wang_attention-based_2022}, is specific to the device used for recording. 
Additionally, own voice transfer characteristics between hearable microphones may differ between devices, e.g., due to the amount of occlusion introduced or the microphone type and placement.
For each new device, a new dataset of recorded own voice signals would be required. 
It is not clear how much recording effort is actually required to achieve satisfactory trained system performance: In e.g.,~\cite{liu_multichannel_2020, yu_time-domain_2020}, recordings of a single talker have been used, while in e.g.,~\cite{wang_multi-modal_2022, wang_fusing_2022}, recordings of several hundred talkers have been used.
In~\cite{wang_fusing_2022}, it was proposed to simulate bone-conduction microphone own voice signals using a DNN-based model in an adversarial training paradigm. 
While the semi-supervised scheme was able to reduce the amount of recorded own voice signals by half without sacrificing trained system performance compared to supervised training with the full dataset, the performance of the trained own voice reconstruction system deteriorates when the amount of recorded signals is reduced further. 

%
Since recording a large amount of own voice signals requires a lot of recording effort, 
it has also been proposed to perform data augmentation by simulating own voice signals to enable training with less recording effort than when using only recorded own voice signals directly as training data~\cite{ohlenbusch_training_2022, he_towards_2023, hauret_configurable_2023, edraki_speaker_2024}. 
In~\cite{ohlenbusch_training_2022}, a speech-independent data augmentation technique has been proposed for training an own voice reconstruction system using simulated in-ear own voice signals.
The data augmentation technique employed a model of own voice transfer characteristics based on RTFs between an outer microphone and an in-ear microphone. 
Trained system performance increased with a higher number of RTFs estimated both from different segments of each talker and from a higher number of talkers. 
The ability of the trained system to reconstruct recorded in-ear own voice signals notably improved by fine-tuning with recorded signals after training with simulated signals. 
In~\cite{he_towards_2023}, a similar speech-independent data augmentation technique was used to simulate bone-conduction accelerometer signals for training an own voice reconstruction system.
In~\cite{hauret_configurable_2023} a speech-independent data augmentation technique has been proposed for training an own voice reconstruction system in which a measured RTF between a close-talk microphone and a body-conduction microphone was used for simulating body-conduction microphone signals.
To introduce additional variance to the simulated own voice signals, random values were added to the RTF during training, which resulted in a performance increase.
In~\cite{edraki_speaker_2024}, it was proposed to perform talker-specific fine-tuning after training on simulated data obtained by a speech-independent augmentation procedure similar to~\cite{ohlenbusch_training_2022, he_towards_2023, hauret_configurable_2023}. A considerable performance gain from fine-tuning was achieved even when only few recorded utterances were available.

%
In this paper, we propose speech-dependent data augmentation techniques for training an own voice reconstruction system by using models of own voice transfer characteristics. 
The own voice transfer characteristic models were previously proposed in~\cite{ohlenbusch_modeling_2023} and can be used to simulate own voice signals at hearable microphones. 
The proposed data augmentation techniques use a small amount of own voice signals recorded with a hearable device to estimate models of own voice transfer characteristics.
These models can then be used to augment a dataset of single-channel speech signals
by simulating in-ear own voice signals that are then used as training data for an own voice reconstruction system.
An experimental evaluation with recorded own voice signals was carried out to investigate the following research questions:
\begin{itemize}
    \item Is speech-dependent data augmentation based on recorded own voice signals beneficial compared to only directly training on recorded own voice signals for training an own voice reconstruction system?
    \item How can recorded own voice signals be used for direct training, for data augmentation, and for fine-tuning to achieve the best own voice reconstruction performance?
    \item How does a decrease in recording effort (in terms of talkers and in terms of utterances per talker) impact own voice reconstruction performance of a system trained using the recorded signals either only directly as training data or in data augmentation and fine-tuning?
\end{itemize}
Results show that speech-dependent data augmentation results in better own voice reconstruction performance than speech-independent data augmentation. 
The proposed speech-dependent individual data augmentation yields better performance compared to only using the recorded own voice signals directly as training dataset. 

%
The remainder of the paper is organized as follows.
In Section~\ref{sec:signal_model}, the considered signal model is described. 
In Section~\ref{sec:transfer_characteristic_models}, own voice transfer characteristic models are presented. 
In Section~\ref{sec:augmentation}, data augmentation techniques for training an own voice reconstruction system based on these models are proposed.
In Section~\ref{sec:evaluation}, The setup of the experimental evaluation for investigating the proposed data augmentation techniques is described.
In Section~\ref{sec:results}, the results of the experimental evaluation are presented and discussed.

\section{Signal model} 
\label{sec:signal_model}

\begin{figure}
    \centering
    \includegraphics{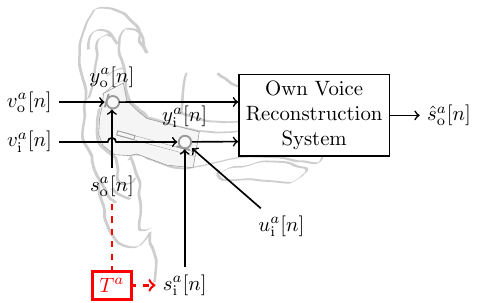}
    \caption{Signal model of noisy multi-microphone own voice reconstruction for a hearable with two microphones (outer, in-ear).
    } 
    \label{fig:sigmodel} 
\end{figure}

Consider a hearable device equipped with an outer microphone and an in-ear microphone, as depicted in Fig.~\ref{fig:sigmodel}.
The signals recorded at each microphone are denoted by subscripts $o$ for the outer and $i$ for the in-ear microphone, respectively. 
We assume that the hearable is worn by a person (referred to as talker) in a noisy environment.
In the time domain, $s_o^a[n]$ and $s_i^a[n]$ denote the own voice of talker $a$ at the outer microphone and at the in-ear microphone, respectively, where $n$ denotes the discrete-time sample index. 
The outer microphone signal $y_o^a[n]$ consists of own voice $s_o^a[n]$ and additive environmental noise $v_o^a[n]$, i.e. 
\begin{equation}
    y_\mathrm{o}^a[n] = s_\mathrm{o}^a[n] + v_\mathrm{o}^a[n].
\end{equation}
Similarly, the in-ear microphone signal $y_i^a[n]$ consists of own voice $s_i^a[n]$ and environmental noise $v_\mathrm{i}^a[n]$ and body-produced noise $u_\mathrm{i}^a[n]$ (e.g., breathing sounds, heartbeats), i.e.
\begin{equation}
    y_\mathrm{i}^a[n] = s_\mathrm{i}^a[n] + v_\mathrm{i}^a[n] + u_\mathrm{i}^a[n].
\end{equation}
The own voice components of talker $a$ at the outer microphone and the in-ear microphone are assumed to be related by the own voice transfer characteristics $T^a\{\cdot\}$, i.e.
\begin{equation}
    s_\mathrm{i}^a[n] = T^a\left\{s_\mathrm{o}^a[n]\right\}.
\end{equation}
In this paper, we further assume that the own voice transfer characteristics $T^a\{\cdot\}$ can be modeled as a time-varying linear system, which is approximated in the short-time Fourier transform (STFT) domain\footnote{
This approximation is only time-varying between STFT frames and not within a single STFT frame, and circular convolutions effects are also neglected in this approximation, but can be reduced by appropriate windowing.} as
\begin{equation}
    \label{eq:stft_approximation}
    S_\mathrm{i}^a(k,l) = H^a(k,l) \cdot S_\mathrm{o}^a(k,l),
\end{equation}
where $k$ denotes the frequency bin index, $l$ denotes the time frame index and $H^a(k,l)$ denotes a time-varying relative transfer function (RTF) of talker $a$ between the microphone at the entrance of the occluded ear canal and the in-ear microphone. 
In the following section, own voice transfer characteristic models based on estimating $H^a(k,l)$ from recorded own voice signals in quiet are presented. 

\section{Own voice transfer characteristic models} 
\label{sec:transfer_characteristic_models}
In this section, several transfer characteristic models are presented that can be used to simulate own voice signals from a dataset of single-channel speech signals.
These models were previously proposed in~\cite{ohlenbusch_modeling_2023}. 
For model estimation, recorded microphone signals of own voice in quiet without environmental noise are used (i.e., assuming $v_o^a[n]=0$ and $v_i^a[n]=0$).

\subsection{Speech-independent models}
\label{sec:speechindepmodel}
If own voice transfer characteristics are assumed speech-independent, the individual transfer characteristics of talker $a$ can be modeled as a linear time-invariant RTF between the outer microphone and the in-ear microphone.
This RTF $\hat{H}^a(k)$ is estimated using the well-known least squares approach~\cite{avargel_multiplicative_2007}.
Considering all STFT frames of the recorded microphone signals from talker $a$ used for estimation, the least-squares RTF estimate is given by
\begin{equation}
\hat{H}^a(k) = \frac{ \sum_l Y_\mathrm{i}^a(k,l) \cdot Y_\mathrm{o}^{a,*}(k,l) }{ \sum_l |Y_\mathrm{o}^a(k,l)|^2},
\end{equation}
where $\cdot^*$ denotes complex conjugation.
Since this model is only based on recorded microphone signals of talker $a$, it is an individual model.

%
A talker-averaged model can be obtained by estimating an RTF over all STFT frames of the recorded microphone signals of all talkers used for estimation instead, i.e.,
\begin{equation} 
    \hat{H}^\mathrm{avg}(k) = \frac{\sum_{a}  \sum_l Y_\mathrm{i}^a(k,l) \cdot Y_\mathrm{o}^{a,*}(k,l)  }{ \sum_{a}  \sum_l |Y_\mathrm{o}^a(k,l)|^2}.
\end{equation}

\subsection{Speech-dependent models} 
\label{sec:speechdepmodel}
Since own voice transfer characteristics likely depend on speech content, it was proposed in~\cite{ohlenbusch_modeling_2023} to model
the transfer characteristics of talker $a$ using a time-varying speech-dependent model.
For this purpose, a frame-wise phoneme sequence $p(l) \in 1,~\dots,~P$ with $P$ possible phoneme classes is obtained from the outer microphone signal $y_\mathrm{o}^a[n]$.
Assuming that the transfer characteristics for each phoneme can be modeled using a (time-invariant) RTF, the RTF for the phoneme $p^\prime$ can be estimated from all frames in which this phoneme is present as
\begin{equation}
    \hat{H}_{p^\prime}^a(k) = \frac
    {\sum_{p(l)=p^\prime}  Y_\mathrm{i}^a(k,l) \cdot Y_\mathrm{o}^{a,*}(k,l)}
    {\sum_{p(l)=p^\prime}  |Y_\mathrm{o}^a(k,l)|^2}.
\end{equation}
Since this model is only based on recorded microphone signals of talker $a$, it is an individual model.

A talker-averaged model can be obtained by estimating an RTF for each phoneme over all STFT frames of the recorded microphone signals of multiple talkers used for estimation in which this phoneme is present instead, i.e.,
\begin{equation}
    \hat{H}_{p^\prime}^\mathrm{avg}(k) = \frac{\sum_{a} \sum_{p(l)=p^\prime} Y_\mathrm{i}^a(k,l) \cdot Y_\mathrm{o}^{a,*}(k,l) }{ \sum_{a}  \sum_{p(l)=p^\prime} |Y_\mathrm{o}^a(k,l)|^2}.
\end{equation}
Although talker-averaged models do not account for individual differences, on average they are able to estimate in-ear own voice signals of different talkers $b\neq a$ better than individual models~\cite{ohlenbusch_modeling_2023}.

\section{Data augmentation techniques using transfer characteristic models}
\label{sec:augmentation}
In this section, we propose data augmentation techniques using the transfer characteristic models presented in Section~\ref{sec:transfer_characteristic_models} for simulating in-ear own voice signals based on single-channel speech signals from a large dataset. 
The models are estimated either from recorded own voice signals of a single talker $a$ (individual models) or from recorded own voice signals of multiple talkers (talker-averaged models). 
The speech signals from the dataset are utilized as outer microphone signals (see Fig.\ref{fig:data_augmentation}) to simulate the corresponding in-ear microphone signals.
For data augmentation, it is assumed that the speech arriving at the microphone originally used for recording the speech signals in the dataset is spectrally similar to the speech arriving at the outer microphone.
Each speech signal from the dataset is assumed to be recorded from a different talker $b$, for whom there is no transfer characteristic model available, so that either an individual model of a different talker or a talker-averaged model needs to be used for augmentation.

\subsection{General augmentation procedure}
In order to obtain an augmented in-ear own voice signal $\hat{s}_i^b[n]$ of talker $b$, a speech signal from the dataset is selected as the outer microphone own voice signal $s_o^b[n]$. 
Augmentation with the own voice transfer characteristic models is performed within a weighted overlap-add (WOLA) scheme. 
First, the outer microphone own voice signal is transformed to the STFT domain to obtain $S_o^b(k,l)$.
Second, $\hat{S}_i^b(k,l)$ is estimated based on~\eqref{eq:stft_approximation}, but using an estimated own voice transfer characteristic model $\hat{H}(k,l)$, i.e.,
\begin{equation} 
    \hat{S}_\mathrm{i}^b(k,l) = \hat{H}(k,l) \cdot S_\mathrm{o}^b(k,l),
\end{equation}
where it is important to stress that the transfer characteristic model is not obtained from talker $b$, since from this talker no recorded in-ear own voice signals are available. 
Depending on the transfer characteristic model, different RTF estimates are used here.
Finally, the augmented own voice signal $\hat{s}_i^b[n]$ is computed by transforming $\hat{S}_i^b(k,l)$ back to the time domain. 

\subsection{Speech-independent augmentation technique}
For augmentation with the speech-independent models, simulated in-ear own voice signals of talker $b$ are obtained by using a speech-independent RTF of one or several different talkers. 
With the speech-independent individual model, the simulated in-ear own voice STFT is computed, i.e.,
\begin{equation}
    \label{eq:sim_speechindep_indiv}
    \hat{S}_\mathrm{i}^b(k,l) = \hat{H}^a(k) \cdot S_\mathrm{o}^b(k,l),
\end{equation}
whereas with the speech-independent talker-averaged model $\hat{H}^\text{avg}(k)$ is used in place of $\hat{H}^a(k)$ in~\eqref{eq:sim_speechindep_indiv}.

\subsection{Speech-dependent augmentation technique}
For augmentation with the speech-dependent models, the phoneme sequence $p^b(l)$ is obtained from the outer microphone signal $y_o^b[n]$ of talker $b$, and then used to select the corresponding phoneme-specific RTF of the own voice transfer characteristics model in each STFT frame. 
To prevent discontinuities in the RTFs during phoneme transitions, recursive smoothing with smoothing constant $\alpha$ is applied, i.e. 
\begin{equation}
    \tilde{H}_{p^b(l)}^a(k) = \alpha \cdot \tilde{H}_{p^b(l-1)}^a(k) + (1-\alpha) \cdot \hat{H}_{p^b(l)}^a(k),
    \label{eq:smoothing_speechdep}
\end{equation}
where $\tilde{H}_{p^b(l)}^a(k)$ is the smoothed RTF of the speech-dependent individual model.
Using the smoothed RTF, the simulated in-ear own voice STFT is then computed as
\begin{equation}
    \hat{S}_\mathrm{i}^b(k,l) =  \tilde{H}_{p^b(l)}^a(k) \cdot S_\mathrm{o}^b(k,l).
    \label{eq:augmentation_speechdep}
\end{equation} 
When using the speech-dependent talker-averaged model, a smoothed talker-averaged RTF $\tilde{H}_{p^b(l)}^\text{avg}(k)$
is obtained from talker-averaged phoneme-specific RTFs $\hat{H}_{p^b(l)}^\text{avg}(k)$ instead of $\hat{H}_{p^b(l)}^a(k)$ in~\eqref{eq:smoothing_speechdep}.
For simulation in~\eqref{eq:augmentation_speechdep}, the smoothed talker-averaged RTF $\tilde{H}_{p^b(l)}^\text{avg}(k)$ is used in place of $\tilde{H}_{p^b(l)}^a(k)$.

\subsection{Random phoneme individual augmentation technique} 
Directly comparing the speech-dependent and speech-independent augmentation techniques does not only compare the influence of introducing speech-dependent behavior, but also the influence of incorporating additional variance to the training dataset by using $P$ phoneme-specific RTFs per talker instead of a single one in the speech-independent models.
To investigate the influence of the additional variance and the speech-dependency separately, an augmentation technique using random RTF selection instead of speech-dependent RTF selection is introduced here.
This technique also uses the RTFs of the speech-dependent individual model, but instead of obtaining a phoneme sequence from $y_o^b[n]$,
the sequence is randomly generated so that for each frame $l$ a random integer $p^\mathrm{rnd}(l) \in 1,\dots,P$ is used independently of speech content in place of $p^b(l)$ in~\eqref{eq:augmentation_speechdep}.
This technique uses the same amount of RTFs as speech-dependent individual augmentation, but applies the RTFs of random phonemes instead of matching phoneme occurrences with the corresponding phoneme RTFs.

\section{Experimental Setup}
\label{sec:evaluation}
This section describes the experimental setup for the evaluation of the proposed data augmentation techniques.
Section~\ref{sec:datasets} details the datasets, recorded own voice signals, and augmented own voice signals, and their use in DNN training and testing.
Section~\ref{sec:trainingdetails} describes the training and testing procedures and hyperparameters.
Section~\ref{sec:dnn_architecture} describes the DNN architecture used in the evaluation.

\subsection{Datasets and recorded signals}
\label{sec:datasets}
In this section, the signals used for training and testing of own voice reconstruction systems are described in this section. 
In Section~\ref{sec:sc_dataset}, the dataset of single-channel speech signals used as input for the augmentation techniques is described.
In Section~\ref{sec:ownvoice_recordings}, recorded own voice signals used either directly in training and validation, for obtaining own voice transfer characteristic models for data augmentation, or for testing are described.
In Section~\ref{sec:augmented_signals}, the augmented own voice signals used in training and validation and details of the augmentation procedure are described. 
In Section~\ref{sec:noise_signals}, the multi-channel environmental noise signals used in training, validation and test are described.

\subsubsection{Dataset of single-channel speech signals}
\label{sec:sc_dataset}
The dataset of single-channel speech signals used as input for own voice data augmentation during training and validation is the German part of the CommonVoice dataset~\cite{ardila_common_2020} (v11.0, only the \textit{validated} subset). 
While the full dataset consists of 1157~hours of speech signals, 10\,\% of the dataset corresponding to 115.7~hours were used in the experimental evaluation. 
Preliminary experiments suggest this amount is sufficient for training an own voice reconstruction system.

\subsubsection{Recorded own voice signals}
\label{sec:ownvoice_recordings}
Recorded own voice signals from 18 native German talkers (5 female, 13 male) are used.
Utterances of 306 sentences from each talker were recorded, corresponding to approximately 25 to 30 minutes of own voice per talker.
The talkers were wearing the closed-vent variant of the Hearpiece~\cite{denk_one-size-fits-all_2019}, an in-the-ear prototype hearable device with multiple microphones. 
The \textit{Concha} microphone of the device was chosen as the outer microphone.
While recording, the talkers were wearing devices in both ears.
For the experimental evaluation, only recorded signals of the left-side device were considered.
The recorded signals are publicly available on Zenodo\footnote{\url{https://zenodo.org/doi/10.5281/zenodo.10844598}}. 
In this paper, the recorded own voice signals are split into disjunct training, validation and test subsets of 12, 2, and 4 talkers, respectively.

\subsubsection{Augmented own voice signals}
\label{sec:augmented_signals}
As mentioned before, the augmentation procedure is carried out as shown in Fig.~\ref{fig:data_augmentation}.
The single-channel speech signals from the dataset described in Section~\ref{sec:sc_dataset} are used as the outer microphone signals. 
The recorded own voice signals from Section~\ref{sec:ownvoice_recordings} are used to estimate own voice transfer characteristic models (Section~\ref{sec:transfer_characteristic_models}).
The own voice transfer characteristic models are then used for augmenting the outer microphone signals (Section~\ref{sec:augmentation}) to obtain simulated in-ear own voice signals.
These datasets can be used for augmented training and validation.
RTFs are estimated using all utterances available for one talker (individual) or all talkers (talker-averaged) in the training subset concatenated in time~\cite{ohlenbusch_modeling_2023}. 
A phoneme recognition system trained on German speech for $P=62$ phoneme classes was used in the speech-dependent augmentation techniques.
If the speech-dependent augmentation techniques encounter a phoneme during simulation for which no RTF is available, a fallback RTF averaged over the other available phoneme RTFs of the corresponding own voice transfer characteristic model.
This happens when only few recorded utterances are available for estimation of own voice transfer characteristics (as considered in Section~\ref{sec:results_recording_effort}).
Since in-ear own voice signals are band-limited, the own voice transfer characteristic models are applied at a sampling frequency of 5\,kHz.
Before estimation and simulation, signals are downsampled. 
The models are applied with an STFT frame length of 25.6\,ms (128 samples) and 50\,\% overlap.
A square-root Hann window is used in WOLA analysis and synthesis.
After simulation, the simulated in-ear own voice signals are upsampled again.

\subsubsection{Environmental noise} 
\label{sec:noise_signals}
The environmental noise used in training, validation, and testing is an individually spatialized version of the single-channel noise dataset from the fifth DNS challenge~\cite{dubey_icassp_2024}.
It consists of approximately 181 hours of environmental noise.
Training an own voice reconstruction system with individually spatialized noise signals was shown to generalize well to recorded noise signals in~\cite{ohlenbusch_multi-microphone_2024}.
To obtain individually spatialized multi-microphone noise signals, individually measured device-specific transfer functions from multiple directions are used.
Head-related transfer functions (HRTFs) between loudspeakers and the device microphones were measured using exponential sweeps from 80\,Hz to 22.05\,kHz with a duration of 3\,s per sweep.
Loudspeakers were positioned in approximately 1.5\,m distance from the talker position and HRTFs were measured from 8 directions in 45°-steps in the horizontal plane.

With a probability of 0.5, either point source noise signals or pseudo-diffuse noise signals are obtained. 
For obtaining a point source noise signal, the noise signal is filtered with the HRTFs for a single loudspeaker.
For obtaining a pseudo-diffuse noise signal, time-shifted copies of the noise signal are filtered with each of the HRTFs for the different directions, and then added.
Then, in both cases, white noise is added to the in-ear environmental noise signal with a random level uniformly distributed in the range of $[-\infty, -60]$\,dB relative to the in-ear environmental noise signal.
This is done to account for the influence of body-produced noise lowering the coherence of the environmental noise between the microphones.

\subsection{Training and evaluation details}
\label{sec:trainingdetails}
The experimental evaluation is carried out at a sampling frequency of 16\,kHz.
STFTs are computed with frame length of 32\,ms (512 samples) and 50\,\% overlap.
A square-root Hann window is used in WOLA analysis and synthesis.
%
During training, utterances are cut to a length of 3\,s.
Own voice signals and environmental noise signals are mixed to a random signal to noise-ratio (SNR) at the outer microphone, uniformly distributed in the range of $[-10, 25]$\,dB. 
Only own voice signals and environmental noise signals of the same talker/device user are mixed, so that they are individually matched.
The in-ear microphone environmental noise signal is scaled by the same amount as the outer microphone environmental noise signal, so that the SNR differences between microphones are preserved.
Next, mean-variance normalization is applied to the noisy signals for each microphone independently. 
The mean and variance of the noisy outer microphone signals are also used to scale the corresponding clean outer microphone own voice signals used as training target by the same amount as the own voice component in the noisy signal~\cite{braun_data_2020}.
During training, a batch consists of 4 utterances. 
The ADAM optimizer~\cite{kingma_adam_2015} is used with an initial learning rate of $10^{-4}$.
For fine-tuning, an initial learning rate of $10^{-5}$ is used.
Training is carried out to a maximum of 100 epochs, where one epoch corresponds to iterating over the entire own voice training subset once.
The learning rate is halved after 3 consecutive epochs without improvement of the validation loss. 
The training is stopped early after 6 consecutive epochs without improvement of the validation loss.
The combined $L_1$ loss in time domain and in frequency domain is used as the loss function~\cite{wang_stft-domain_2023}.  
The PyTorch~\cite{paszke_pytorch_2019} (v1.10) framework was used for implementation, and the computations were executed on two NVIDIA GeForce RTX 2080 SUPER GPUs.

%
Five different augmentation techniques described in Section~\ref{sec:augmentation} are evaluated: 
Speech-independent individual augmentation, speech-independent talker-averaged augmentation, speech-dependent individual augmentation, speech-dependent talker-averaged augmentation, and random phoneme individual augmentation. 
The augmentation techniques are not used in test.
To compare the trained system performance with and without data augmentation, training using only the recorded own voice signals directly as training or validation input is evaluated as well.
%
For testing, evaluation metrics were applied to the estimated outer microphone own voice signal, and the clean outer microphone own voice signal is used as the reference signal. 
The evaluation metrics are wideband perceptual evaluation of speech quality (PESQ)~\cite{international_telecommunications_union_itu_itu-t_2001} and short-time objective intelligibility (STOI)~\cite{taal_algorithm_2011}.
An increase in each metric indicates performance improvement.
For testing, SNRs of $[-10,-5,0,5,10]$\,dB are considered. 
In Section~\ref{sec:results_recording_effort}, when reducing the number of talkers, only randomly selected fractions of the talkers in the training subset are used in the proposed individual speech-dependent augmentation technique, while for each talker all utterances are used. 
The considered numbers of talkers are $[1,2,3,4,6,8,10,12]$.
When reducing the number of utterances per talker, only randomly selected fractions of the utterances per talker in the training subset are used in the proposed individual speech-dependent augmentation technique, while recorded signals of all talkers are used. The considered number of utterances per talker are $[1,3,6,12,25,75,150,306]$.

\subsection{DNN architecture} 
\label{sec:dnn_architecture}
\begin{figure*}
    \centering
    \includegraphics{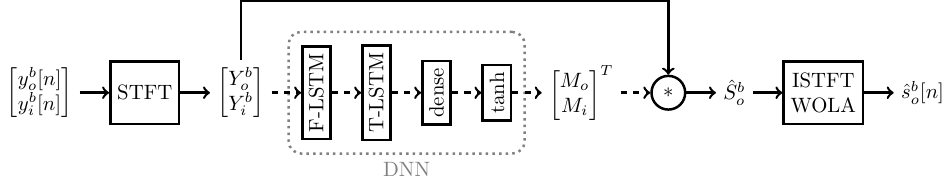} 
    \caption{Modified FT-JNF architecture based on~\cite{tesch_insights_2023} used as own voice reconstruction system. STFT indices are omitted.}
    \label{fig:dnn_architecture}
\end{figure*}
Fig.~\ref{fig:dnn_architecture} depicts the DNN architecture used in the experiments. 
It is based on the FT-JNF architecture proposed in~\cite{tesch_insights_2023}.
The noisy microphone signals are transformed into the STFT domain before being processed by the DNN.
The inputs to the DNN are the real and imaginary part of the complex-valued noisy outer microphone STFT and in-ear microphone STFT, respectively.  
The DNN architecture consists of a uni-directional LSTM layer with 512 hidden units operating along the frequency dimension (F-LSTM), then a uni-directional LSTM layer with 128 hidden units operating along the time dimension (T-LSTM), followed by a dense layer combining the 128 outputs of the second LSTM layer to 4 outputs. 
The dense layer is followed by tanh activation.  
The outputs of the DNN are the real and imaginary part of two complex-valued masks $M_o(k,l)$ and $M_i(k,l)$. This is different from~\cite{tesch_insights_2023}, where only one mask is estimated, and was found to lead to better performance for own voice reconstruction compared to only estimating a single mask in~\cite{ohlenbusch_multi-microphone_2024}.
After obtaining the complex-valued masks, the estimated own voice signal 
of talker $b$ at the outer microphone is computed as
\begin{equation}
    \hat{S}_o^b(k,l) = 
    \sum_{ \mathclap{m\in\{o, i\}} } 
        M_m(k,l) \cdot Y_m^b(k,l).
        \label{eq:ftjnf_masking}
\end{equation}
Since the training is carried out for multiple talkers, the masks are not specific to the target talker $b$, meaning the approach to own voice reconstruction in this paper is non-individualized.
After computing~\eqref{eq:ftjnf_masking}, the inverse STFT is computed and a weighted overlap-add (WOLA) scheme is employed to obtain the time-domain signal $\hat{s}_o^b[n]$.
Overall, the DNN consists of approximately 1.4\,M parameters.
The algorithmic latency of the DNN architecture corresponds to the STFT frame shift of 16\,ms.


\section{Results and Discussion} 
\label{sec:results}
In this section, the results of the experimental evaluation are presented and discussed. 
In Section~\ref{sec:results_augmentation}, the data augmentation techniques proposed in Section~\ref{sec:augmentation} are compared to training only using recorded signals in terms of trained system performance for own voice reconstruction. 
This is done to investigate whether the data augmentation techniques using recorded own voice signals for estimating transfer characteristic models are beneficial compared to only training on the recorded own voice signals directly.  
In Section~\ref{sec:results_finetuning}, additional performance gains from fine-tuning different layers of the DNN architecture previously trained using speech-dependent individual augmentation are investigated.
In Section~\ref{sec:results_recording_effort}, the trade-off between recording effort and trained system performance for direct training, training only on augmented data, and fine-tuning is explored. 

\subsection{Training using augmented own voice signals} 
\label{sec:results_augmentation}
\begin{figure} 
    \centering
\includegraphics[width=.5\textwidth]{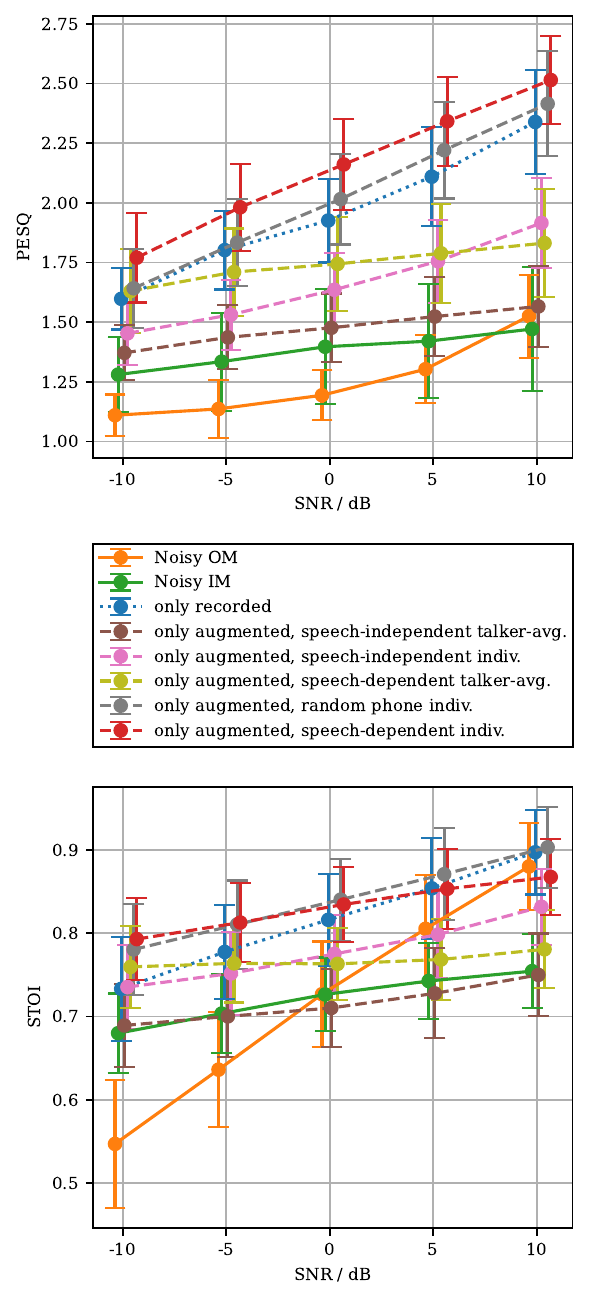}
    \caption{Own voice reconstruction performance
    of training only on recorded signals directly and training with data augmentation techniques. The results for the unprocessed microphone signals are also shown.
    The results are averaged over the test subset. 
	Error bars indicate standard deviations over all test subset examples. 
	Data points are slightly shifted horizontally to improve readability.} 
    \label{fig:speech_results} 
\end{figure}

Fig.~\ref{fig:speech_results} shows the trained system performance in terms of PESQ and STOI for DNNs trained using speech-independent and speech-dependent data augmentation techniques (individual and talker-averaged), and a DNN trained using only recorded signals. The results for the unprocessed noisy outer and in-ear microphone signals are also shown.
Speech-dependent augmentation leads to a consistent improvement over speech-independent augmentation for both individual and talker-averaged augmentation.
It can be observed that speech-dependent individual augmentation leads to the highest performance in terms of both metrics. 
In particular, this augmentation technique leads to better results than training a DNN only on the recorded own voice signals. 
Additionally, individual augmentation achieves better performance than talker-averaged augmentation. 

%
Individual random phoneme augmentation leads to higher trained system performance than speech-independent individual augmentation.
This increase in performance is due to the additional variance in the augmented training data from augmenting with $P$ instead of a single RTF per talker.
It should be noted that this observation matches previously reported results~\cite{ohlenbusch_training_2022, hauret_configurable_2023}, where an increase in the number of RTFs or introducing additional variance in augmentation lead to increased trained system performance.
When comparing individual random phoneme augmentation with speech-dependent individual augmentation, it is observed that using phoneme-matched RTF selection instead of random selection leads to better results in terms of PESQ.
This indicates that speech-dependent modeling yields a benefit in addition to the performance gained from additional variance in the training data. 

%
In summary, the results show that the proposed speech-dependent individual augmentation technique is able to simulate own voice signals sufficiently well for training a DNN to reconstruct own voice from recorded own voice signals.
In the following sections, only speech-dependent individual augmentation is investigated further.

\subsection{Use of recorded signals for additional fine-tuning}
\label{sec:results_finetuning}
\begin{figure}
    \centering
    \includegraphics[width=.5\textwidth]{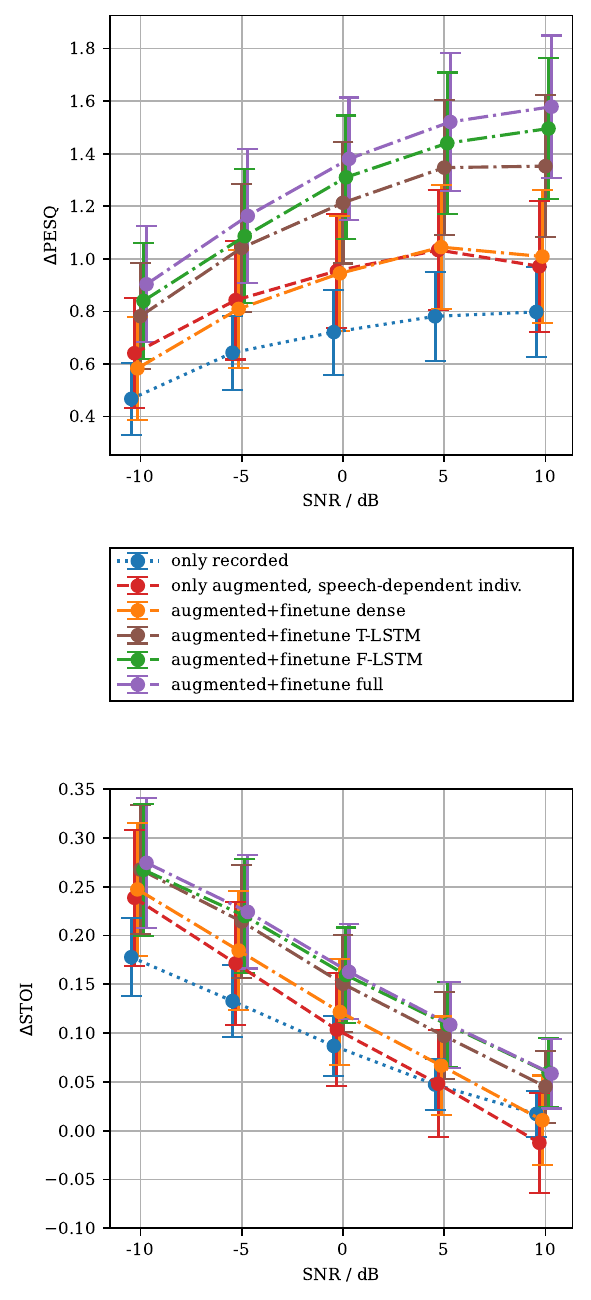}
    \caption{
    Own voice reconstruction performance 
    for a DNN trained only on recorded own voice signals, 
    training only on augmented own voice signals obtained with speech-dependent individual augmentation,
    and training first with speech-dependent individual augmentation and then fine-tuning different parts of the DNN on the recorded own voice signals.
    Results are averaged over the test subset. 
	Error bars indicate standard deviations over all test subset examples.
	Data points are slightly shifted horizontally to improve readability.
    } 
    \label{fig:results_finetuning}
\end{figure}

Fig.~\ref{fig:results_finetuning} shows performance gains in terms of PESQ and STOI from fine-tuning different layers of the DNN previously trained using speech-dependent individual augmentation.
Different from the previous results, the improvements ($\Delta$) over the noisy outer microphone signals are shown to facilitate easier comparisons. 
The improvements in PESQ are larger in higher SNRs, while the improvements in STOI are larger in lower SNRs. 
All fine-tuning approaches lead to equal or better results than only training on augmented own voice signals and better results than only training on recorded signals.
When only the dense layer is fine-tuned on the recorded own voice signals after training on augmented own voice signals, there is barely any change in PESQ and only small improvements in STOI compared to training on augmented own voice signals. 
When only the T-LSTM layer or only the F-LSTM layer are fine-tuned, the improvement compared to training only on augmented own voice signals is larger than compared to fine-tuning the dense layer. 
Fine-tuning the F-LSTM layer leads to slightly better performance than fine-tuning the T-LSTM layer. 
When the full DNN is fine-tuned after augmented training, the best results are obtained. 

%
In summary, the results show that fine-tuning on recorded own voice signals is beneficial in addition to augmented training using transfer characteristic models obtained from the recorded own voice signals.
In the following section, only fine-tuning the full DNN is investigated further.

\subsection{Trade-off between performance and recording effort}
\label{sec:results_recording_effort}
The trade-off between own voice reconstruction performance and recording effort for training an own voice reconstruction system is investigated by separately reducing the number of talkers and the number of utterances per talker.
Three training conditions are compared: Training the DNN only on recorded own voice signals, training the DNN only on augmented own voice signals, and first training the DNN on augmented own voice signals and then fine-tuning on recorded own voice signals.
\begin{figure*} 
    \centering 
    \subfloat[PESQ improvement over noisy outer microphone signals when reducing the number of talkers.\label{fig:talkercount_PESQ}]{%
    \includegraphics[width=0.5\textwidth]{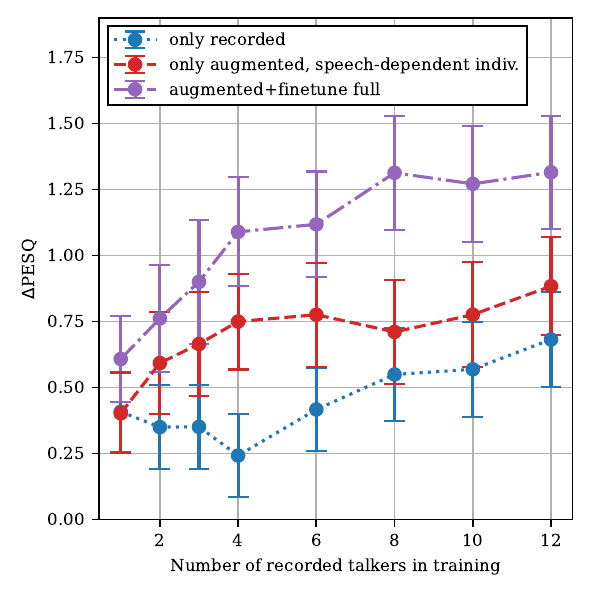}}
    \subfloat[PESQ improvement over noisy outer microphone signals when reducing the number of utterances per talker.\label{fig:uttamount_PESQ}]{%
    \includegraphics[width=0.5\textwidth]{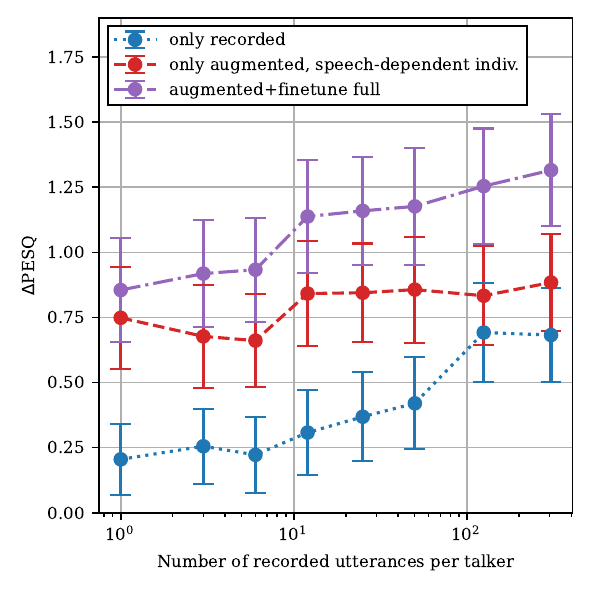}}
    \\
    \subfloat[STOI improvement over noisy outer microphone signals when reducing the number of talkers.\label{fig:talkercount_STOI}]{%
    \includegraphics[width=0.5\textwidth]{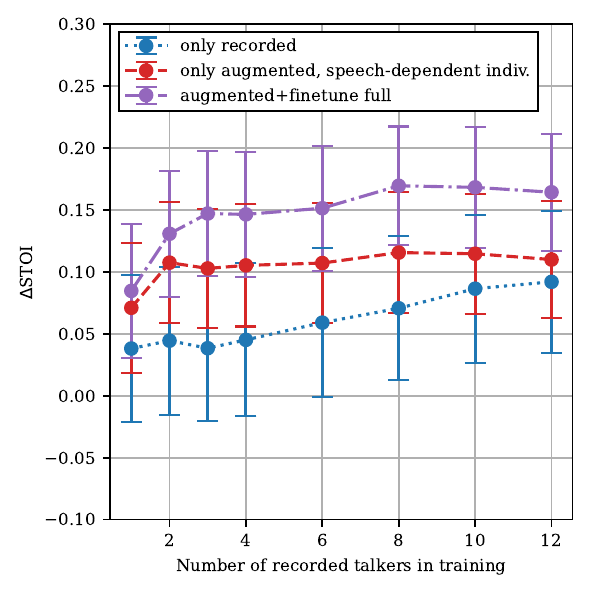}}
    \subfloat[STOI improvement over noisy outer microphone signals when reducing the number of utterances per talker.\label{fig:uttamount_STOI}]{%
    \includegraphics[width=0.5\textwidth]{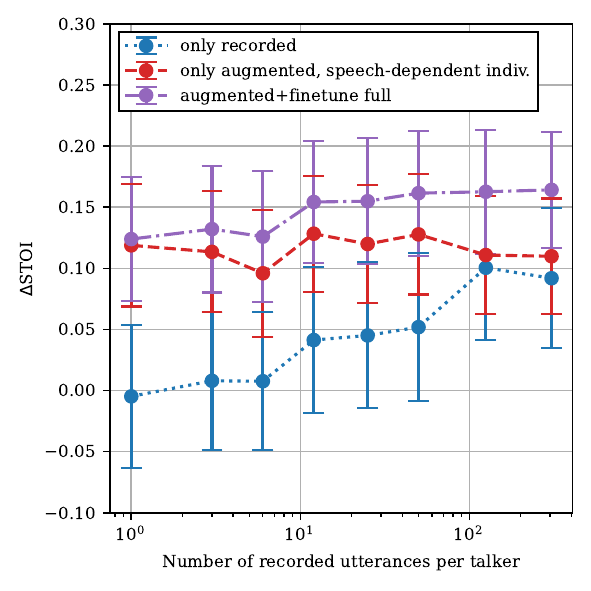}}
    \caption{
    Own voice reconstruction performance with a reduced number of talkers~\protect\subref{fig:talkercount_PESQ},~\protect\subref{fig:talkercount_STOI}, 
    and a reduced number of utterances per talker~\protect\subref{fig:uttamount_PESQ},~\protect\subref{fig:uttamount_STOI} 
    available for training. 
	Results are averaged over the test subset and SNRs. 
	Error bars indicate standard deviations over all test subset examples and SNRs.
    } 
    \label{fig:results_recording_effort}
\end{figure*}

Fig.~\ref{fig:results_recording_effort} shows the $\Delta$PESQ and $\Delta$STOI scores for DNNs trained in the three training conditions.
For both the number of talkers and utterances per talker, first training on augmented own voice signals and then fine-tuning on recorded own voice signals leads to higher performance compared to only training on augmented own voice signals and also compared to training only on recorded own voice signals. 
In most cases, only training on augmented own voice signals leads to higher performance than training only on recorded own voice signals.

%
In terms of number of talkers, the performance achieved by all three training conditions 
is reduced when the number of talkers is reduced, particularly for less than 8 talkers. 
The improvement achieved by fine-tuning over only augmented training is larger when recorded own voice signals of more talkers are used. 
The difference between training only on recorded own voice signals and training only on augmented own voice signals is larger for fewer talkers, except for a single talker, where the performance in PESQ is almost the same and the difference in STOI is small. 
These results suggest that the augmentation is particularly effective when only recorded own voice signals of few talkers are available, and additional fine-tuning is more effective when recorded own voice signals of more talkers are available.
The results show a large increase in terms of PESQ between a single talker and 6 to 8 talkers, suggesting that the proposed augmentation technique is more effective when recorded own voice signals of multiple talkers are available.
Adding more talkers for estimation of transfer characteristic models as well as for fine-tuning improves the trained system performance. 

%
In terms of number of utterances per talker, the performance achieved by training a DNN only on recorded own voice signals strongly decreases if a lower number of utterances is available. 
When training only on augmented own voice signals, the performance only slightly decreases when the number of utterances is reduced.
Even with only few utterances available, there is still a large improvement through processing by the DNN in both metrics. 
When fine-tuning on recorded own voice signals after training on augmented own voice signals, there is a small benefit compared to only training on augmented own voice signals with few utterances, but a larger benefit compared to only training on augmented own voice signals with more utterances available. 
Similar to the results for the number of talkers, the results show that augmented training with the proposed augmentation technique achieves the largest benefit compared to training only on recorded own voice signals when a low number of utterances is available.
Fine-tuning is more effective when more recorded utterances per talker are available.
Since the trained system performance for low numbers of utterances is still relatively high, it appears that adding more recorded utterances per talker is less beneficial for using the proposed augmentation technique than adding more recorded own voice signals of different talkers.

%
In summary, the results show that the proposed speech-dependent augmentation technique enables to train an own voice reconstruction system even when few recorded own voice signals are available. 
Slightly increasing the recording effort can lead to performance gains, especially in the case of few recorded talkers.

\section{Conclusion}

In this paper, we have proposed speech-dependent own voice data augmentation techniques for training DNN-based own voice reconstruction systems. 
The techniques use own voice transfer characteristic models estimated from few available recorded own voice signals, and can be applied to large dataset of single-channel speech signals to simulate large amounts of own voice signals as DNN training data.
It was investigated whether accounting for speech-dependency and individual differences in own voice transfer characteristic models used for data augmentation is beneficial for trained system performance. 
Experimental results show that speech-dependent individual data augmentation leads to better performance compared to other augmentation techniques, as well as to training only on the available recorded own voice signals.
The proposed speech-dependent individual augmentation technique can exploit the availability of single-channel speech data for increasing trained system performance without increasing the amount of recorded own voice signals available for training. 
Experimental results also show that fine-tuning on recorded own voice signals after training with the proposed speech-dependent individual data augmentation can increase performance even further, even when few recorded signals are available.  
Additionally, it was investigated how much the number of talkers and the number of utterances per talker in recorded own voice signals influences the performance of systems trained using the proposed augmentation techniques. 
It was found that the influence of the number of talkers is rather high, while the influence of the number of utterances per talker is smaller. 
It is concluded that the proposed speech-dependent individual augmentation technique is highly beneficial for training an own voice reconstruction system.


\bibliographystyle{IEEEtran} 
\bibliography{IEEEabrv, bibcontrol, manual_bib}

\end{document}